# Temperature behavior of the Optical Absorption and Photoluminescence Spectra of InP/ZnS Nanocrystals with a Stabilizing Polyvinylpyrrolidone-based Coating

S.S. Savchenko[1], A.S. Vokhmintsev[1], M.S. Karabanalov[1],
A.M.A. Henaish[1,2], I.A. Weinstein[1,3]*

[1] NANOTECH Center, Ural Federal University,
Ekaterinburg, Mira street 19, Russia, 620002

[2] Department of Physics, Tanta University,
Tanta, Egypt, 31527

[3] Institute of Metallurgy, Ural Branch of Russian Academy of Sciences,
Ekaterinburg, Amundsena street 101, Russia, 620016

*e-mail: i.a.weinstein@urfu.ru

**Abstract:** The present paper deals with the results of a research work on the optical properties of colloidal InP/ZnS nanocrystals stabilized with a heterobifunctional polyvinylpyrrolidone polymer. We have analyzed the absorption and photoluminescence spectra of the samples as solutions with various concentrations and deposited films, as well as the patterns of their temperature changes in the range of 6.5–296 K. An inhomogeneous broadening of exciton optical bands has been observed to be related to a wide distribution of nanocrystals in size. A temperature shift of the exciton absorption and emission maxima has been found to be due to the interaction with acoustic phonons. It has been shown that the quenching of defect-related luminescence is carried out through local energy levels of dangling bonds of phosphorus atoms involved at the core-shell interface, and the temperature stability of exciton emission is determined by the thickness of the ZnS shell.

## 1. Introduction

Quantum dots (QDs) or semiconductor nanocrystals are zero-dimensional objects and have properties that can be radically distinguished from those of related bulk materials of similar composition. The energy structure of QDs is strongly size-dependent. This is explained by the manifestation of the quantum confinement effect in all three spatial dimensions [1–5]. Currently, the possibilities of applying nanocrystals to design light-emitting devices, laser sources, photocells, phosphors, biotags, sensors, etc. are being intensively investigated [6–9]. Colloidal core/shell III–V compounds-based QDs as an alternative to nanosized cadmium-containing structures with low toxicity and higher photostability are of particular interest [10,11]. At the same time, QDs with an InP core and a ZnS shell are among the most high-potential and gained widespread acceptance in practice.

Understanding of the peculiarities of forming the mechanisms of exciton absorption and luminescence of InP/ZnS QDs in a wide temperature range will make it possible to obtain information, including practice-oriented one, for a feasible control of the synthesis regimes of such systems with a high quantum yield and an optical response stable to external conditions.

The InP/ZnS structures belong to type I systems having the core bandgap located within the appropriate band of the shell. The latter serves as a barrier between the surface of the optically active core and the surrounding space to enhance the luminescence quantum yield and resistance to photodegradation, with the sensitivity-to-change properties weakening in the local surrounding. As the core changes in the size, the emission of InP/ZnS varies in the visible and near-IR regions. Simultaneously, the quantum yield of some samples becomes comparable to the most efficient cadmium chalcogenides-based nanocrystals. For reducing the toxicity of QDs, a method of layer-by-layer growing of the ZnS shell is widely used for nanocrystals containing cadmium, lead, or mercury without succeeding their complete harmlessness [12,13]. It is known that the biocompatibility of InP is much higher [14]. Whilst, the phototoxicity inherent in all nanocrystals and being due to the generation of reactive oxygen species is retained. The rate of this process was found to be inversely proportional to the number of ZnS layers on the core surface [15].

A comparison of the harmful impact of InP/ZnS nanocrystals on various cell cultures, bacteria, and animals showed that they are much safer than cadmium-containing water-soluble those in a ZnS shell [14,16,17]. The reason for this is the lower mobility of indium, as compared to Cd ions that are capable of releasing from the core despite the existence of a shell. The InP/ZnS system as an addressable optical probe was successfully employed for labeling pancreatic cancer cells [18], of fluorescently detecting adenosine triphosphoric acid [19], of monitoring food freshness [20], etc. Thus, these biocompatible and environmentally safe core/shell QDs can be unlimitedly utilized in any applications. The present paper covers the outcomes of studying the optical properties of colloidal InP/ZnS nanocrystals stabilized with a polyvinylpyrrolidone-based coating in a wide temperature range.

## 2. Results

We examined two ensembles of InP/ZnS colloidal quantum dots with an average core size of 2.0 (S1) and 2.6 (S2) nm. Figure 1 includes images of sample S1, scanned with a JEOL JEM-2100 transmission electron microscope. It can be seen that the particles have a close-to-spherical shape. An atomic structure is resolved for some QDs to indicate their crystalline nature. The interplanar spacing values residing in the range of 0.32–0.34 nm evidence the orientation of particles in the $\langle 111 \rangle$ direction and are consistent with the data of independent works on InP/ZnS synthesized by various methods [21–24].

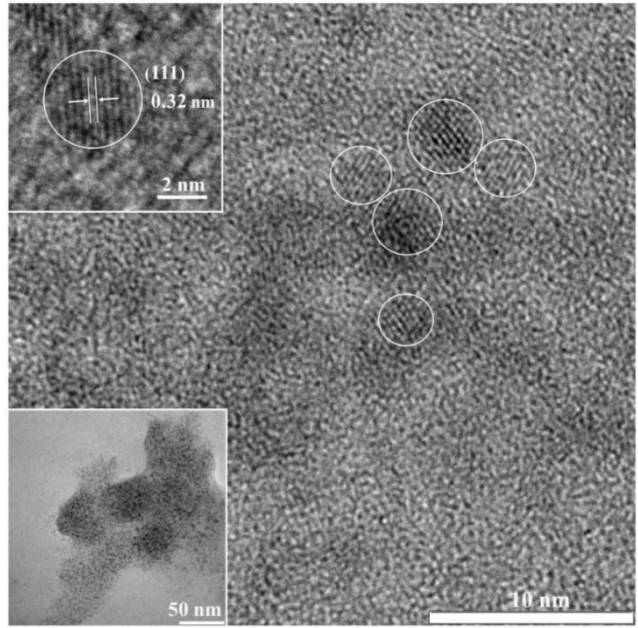

**Figure 1.** TEM image of sample S1 at various scales.

Figure 2 shows the $D$-$\lambda$ dependencies, where $D$ is the optical density and $\lambda$ is the wavelength, for quantum dots of S1 and S2 in the form of colloidal solutions of various concentrations and films obtained by deposition of the appropriate volumes of the initial solution (see the caption to Figure 2). For both solutions and films, two shoulders can be observed in the ranges of 420–480 and 290–320 nm for S1, as well as in the ranges of 530–590 and 300–330 nm for S2. In the case of the first sample, curves 4 and 9 overlap.

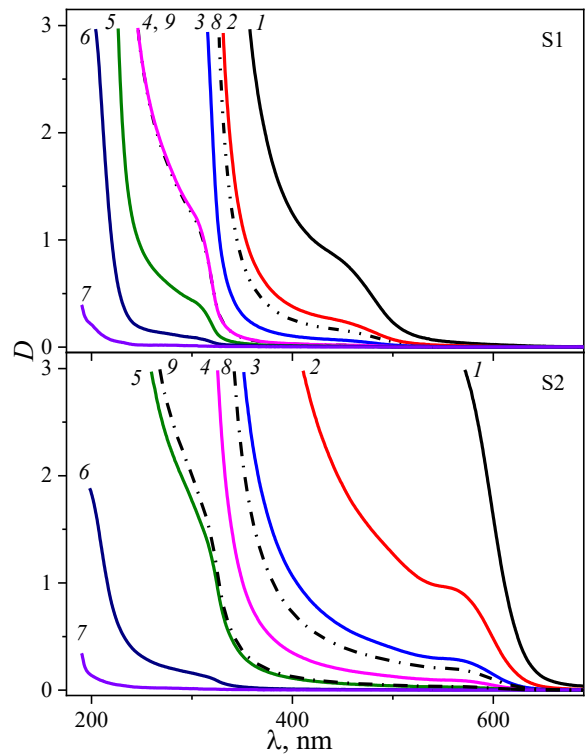

**Figure 2.** OA spectra of InP/ZnS colloidal solutions of various concentrations and deposited QD films. Solutions: (*1*) - 100 g/l, (*2*) - 25 g/l, (*3*) - 6.25 g/l, (*4*) - 1.56 g/l, (*5*) - 390 mg/l, (*6*) - 24 mg/l, (*7*) - 0.1 mg/l. Films: (*8*) - 100 μl, (*9*) - 5 μl.

The temperature-dependence of absorption for each of the samples in the spectral range corresponding to the first shoulder is presented in Figure 3. Upon cooling, a shift of these features occurs towards the short-wavelength side, with the corresponding optical density increasing.

Figure 4 displays the dependencies of the PL intensity $I$ of InP/ZnS nanocrystals studied on the photon energy $E$. At room temperature, a single $E_x$ maximum and asymmetric-shaped emission bands are seen. The plot is characterized by a flatter and extended low-energy part. Upon cooling up to 6.5K, along with an increase in the luminescence intensity over the entire spectrum, this spectral region reveals a pronounced $E_d$ component.

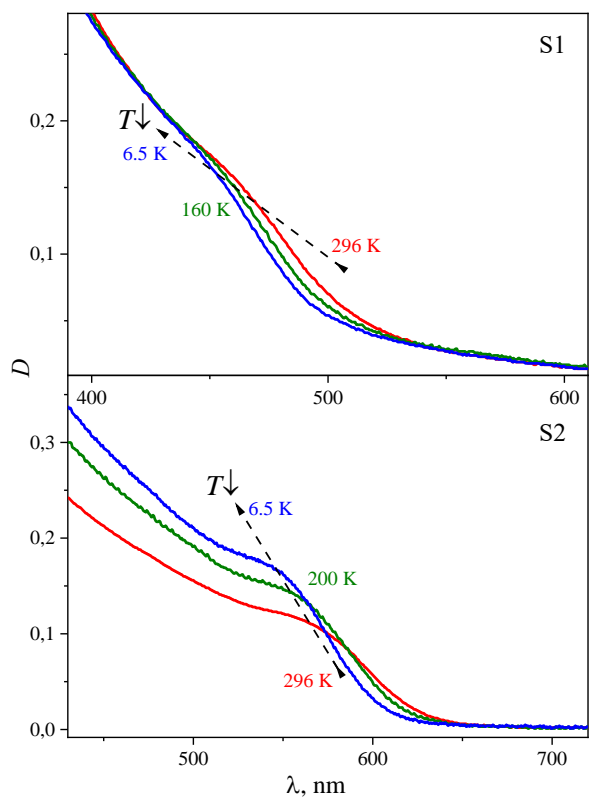

**Figure 3.** InP/ZnS OD spectra measured at different temperatures.

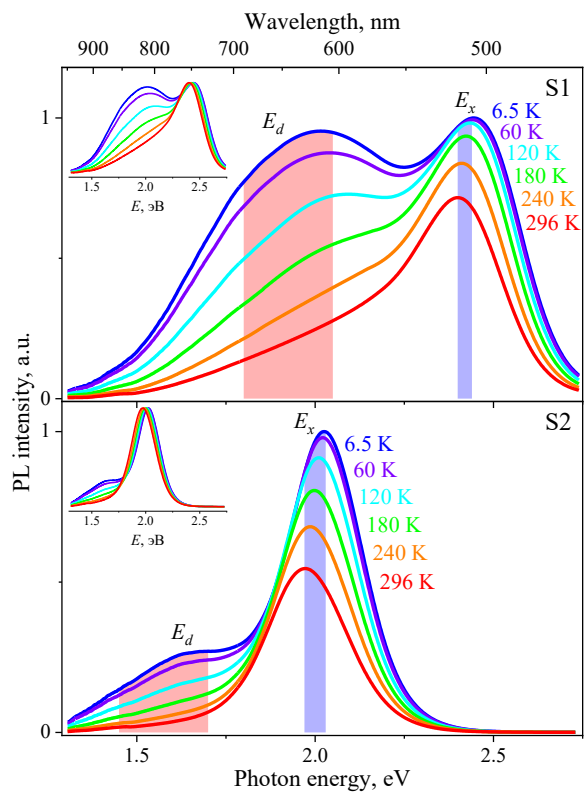

**Figure 4.** InP/ZnS PL spectra measured at different temperatures.

### 3. Discussion

#### 3.1 Optical transitions in the core and shell

An idealized theoretical description predicts a linear, atom-like character of QDs absorption [1, 4, 5]. In reality, the samples demonstrate a continuous absorption spectrum, including maxima and/or shoulders. This is elucidated by the fact that the bands have a certain width, overlap with each other and are distorted due to a background. In this regard, to characterize optical transitions and determine the positions of absorption bands, we resorted to analytical approaches of derivative spectrophotometry [25–27].

The OA spectra contain two optical transitions whose energies are the same for QD samples as both solutions and films (see Figure 5). For S1, the energies $E_1$ and $E_2$ are 2.70 and 3.98 eV; for

S2, they amount to 2.15 and 3.89 eV, respectively. It has previously been found [26, 28] that, for InP/ZnS nanocrystals with an average core size of 2.1 and 2.3 nm, synthesized by an identical technique but coated with a modified polyacrylic acid (PAA), high-energy transitions in deposited films evince themselves as a redshift relative to similar values for solutions. This fact appears to appertain to the effects of the interaction between nanocrystals since the deposition changes the distance between them and can lead to a modification of their electronic structure [29]. In this case, core/shell QDs stabilized with modified PAA [26] are assumed to form more densely packed films with a smaller distance between individual nanocrystals, as compared to PTVP-coated QDs.

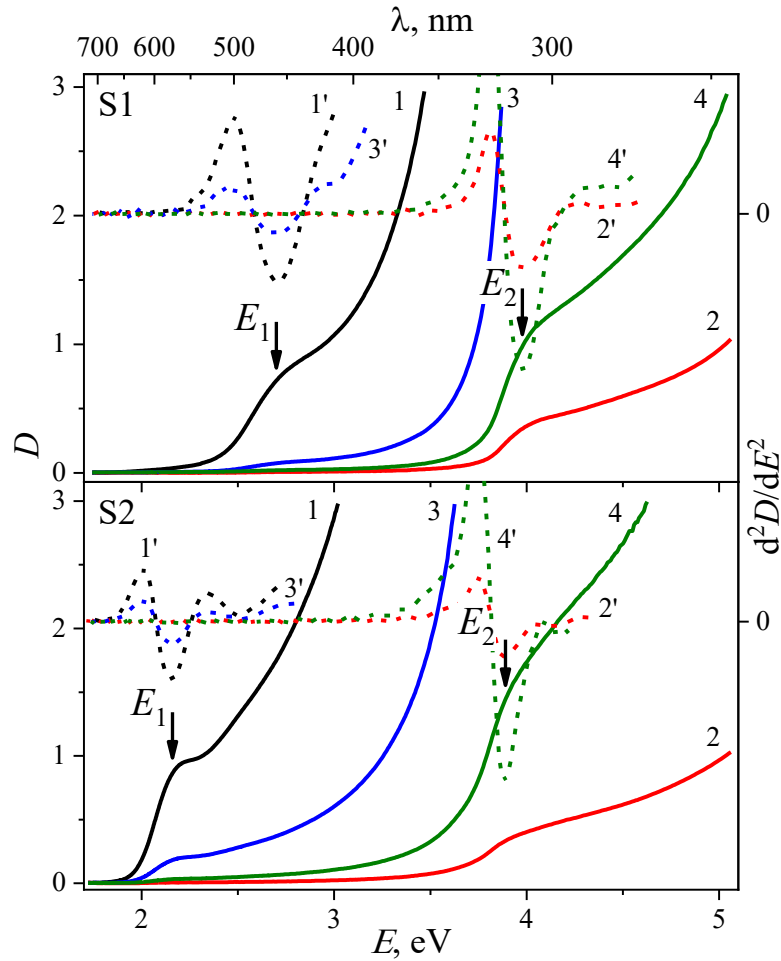

**Figure 5.** OA spectra (solid lines) and their second derivatives (dashed lines) for solutions and films. S1: 1 – solution $c$=100 g/l, 2 – solution $c$=0.39 g/l, 3 and 4 – films. S2: 1 – solution $c$=25 g/l, 2 – solution $c$=0.1 g/l, 3 and 4 – films. The arrows indicate the energies of optical transitions.

The $E_1$ transition is ascribed to the first exciton absorption band of the indium phosphide core [30,31]. In modern scientific literature, this energy is regarded as the bandgap width $E_g$ or the gap between the states corresponding to the top of the valence band and the bottom of the conduction band. The energy $E_2$ exceeds $E_g$ = 3.6 eV for bulk ZnS at $T$ = 290K [32], which may be due to the manifestation of the quantum size effect in the shell of the samples under study. The assumption made above is confirmed by the known findings on the synthesis of ZnS nanopowders and colloids with $E_g$ values in the energy ranges of 3.71 – 4.3 [33,34] and 4.47 – 4.82 eV [35]. The $E_2$ value for the nanocrystals tested is less than that for the PAA-coated InP/ZnS QDs (4.10 eV [26]), which says about a thicker shell of the nanocrystals at hand.

In addition, the ratio of the optical density in the $E_2$ and $E_1$ bands for the studied ensembles is 75 for S1 and 83 for S2 and 5–10 times higher than the appropriate values for PAA-coated QDs previously investigated [36]. On the one hand, a different nature of the bands takes place. On the other hand, the difference observed is also related to the thicker ZnS shell in the QD ensembles. Bearing in mind the above facts, we can conclude that the E2 band is associated with transitions in the ZnS shell and the observable absorption is the sum of the spectra of two semiconductor materials under size quantization conditions. For InP, the discussed region is a short-wavelength one and is characterized by a quasi-continuous spectrum. For ZnS, this range can be regarded as a long wavelength one since $E_g$(ZnS) > $E_g$(InP). Thus, the first exciton absorption band of ZnS is superimposed on the quasi-continuous spectrum of InP.

## 3.2 $E_1$ band offset

To evaluate the spectrum energies $E_1$ at the temperatures of 6.5–296 K, we exploited the derivative spectrophotometry method. In Figure 6, square symbols represent the energies of the first exciton absorption band at the appropriate temperatures. It can be seen that $E_1$ goes up with declining $T$ and behaves as a temperature-induced change in the optical gap width, inherent in semiconductor crystals [32,37]. The solid lines guide the approximation of the experimental data using the Fan relation [27,31,38]:

$$E_1(T) = E_1(0) - A_F \langle n_s \rangle, \langle n_s \rangle = \left[\exp(\hbar\omega / kT) - 1\right]^{-1}. \qquad (1)$$

Here $E_1(0)$ is the optical transition energy at 0 K, eV; $A_F$ is Fan parameter depending on the microscopic properties of the material, eV; $\langle n_s \rangle$ is the Bose-Einstein factor for phonons with energy $\hbar\omega$; $k$ is the Boltzmann constant, eV/K. The dashed lines in Figure 6 offer a description of the experimental data under the assumption of the linear model $E_1(T) = E_1(0) - \beta T$ that is of practical interest. The coefficients β (red circle symbols) for the corresponding temperature ranges are determined. It can be seen that the dependency $E_1(T)$ deviates significantly from the linear one. In addition, Figure 6 contains the temperature curves $dE_1/dT$ (dotted lines) that characterize the temperature function β(T). For its high-temperature limit, the values of the coefficient $β_\infty$ are also marked (red square symbols) [37]. The resulting outcomes are listed in Table 1 in comparison with known data.

The values of the effective phonon energy $\hbar\omega$, Fan parameter $A_F$ and the Huang-Rhys $S$ factor, obtained for the samples S1 and S2, describe the temperature-behavior of the energy $E_1$. They quite meet the values for OA of PAA-coating InP/ZnS [27,36], for the bulk modification of indium phosphide [39], for InP nanowires [40], and for estimates of similar values on the PL spectra in InP/ZnS QDs [41,42]. Thus, the data secured confirm that the temperature shift of the first exciton absorption band is caused by the interaction between excitons and longitudinal acoustic phonon modes.

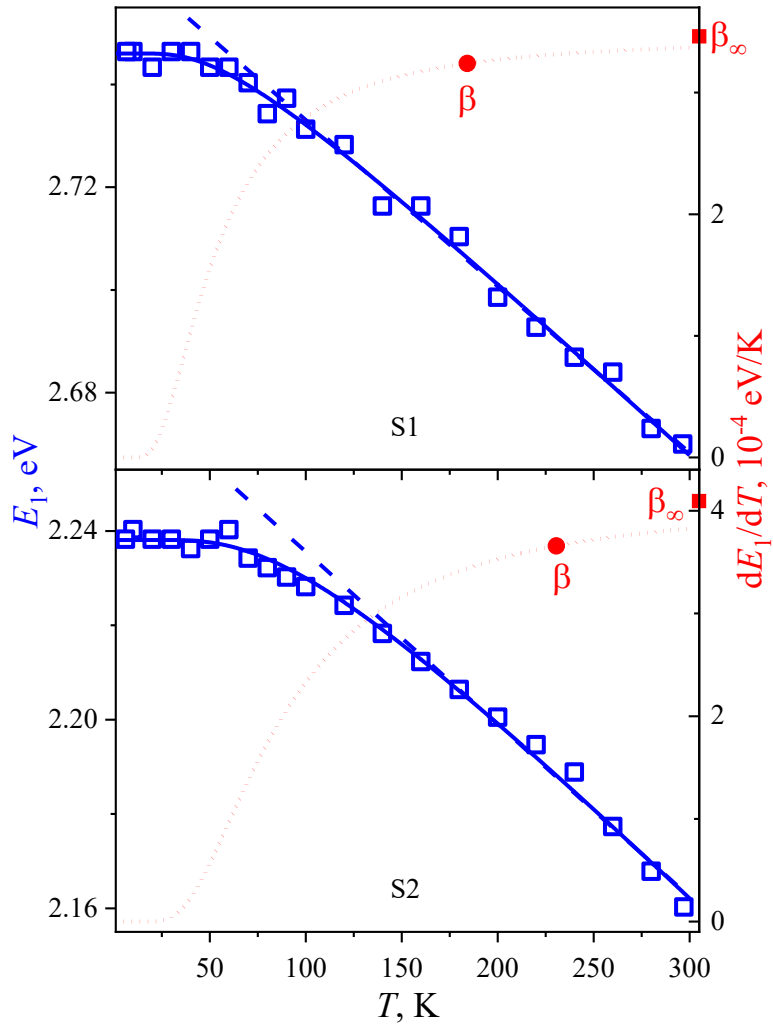

**Figure 6.** Temperature dependence of the $E_1$ transition for InP/ZnS samples.

**Table 1.** Parameters of the temperature shift of the exciton band of OA in QDs.

| Size, nm (coating) | $E_1(0)$, eV | $A_F$, eV | $\hbar\omega$, meV | $S$ | $\beta$, $10^{-4}$ eV/K | $\beta_\infty$, $10^{-4}$ eV/K |
|---|---|---|---|---|---|---|
| 2.0 (S1) (PTVP) | 2.75 | 0.06 | 14<br>12 [41] | 2.00<br>1.95 [41] | 3.24<br>(90-296) | 3.45 |
| 2.6 (S2) (PTVP) | 2.24 | 0.11 | 23<br>23 [42] | 2.37<br>2.45 [42] | 3.66<br>(140-296) | 4.09 |
| 2.1 [27] (PAA) | 2.72 | 0.09 | 15 | 2.98 | 4.76<br>(90-296) | 5.13 |
| 2.3 [36] (PAA) | 2.49 | 0.21 | 31 | 3.49 | 5.36<br>(180-330) | 6.01 |
| InP nanowires [40] | 1.50 | 0.09* | 20.5 | 2.20* | – | 3.78* |
| Bulk InP [37,39] | 1.42 | 0.05 | 14 | 1.78* | 2.9 | 3.09 |

* our estimate based on the aforecited works.

According to Table 1, the size effect manifests itself not only through the absolute value of the transition energy but also through its temperature behavior. Comparing different groups of QDs with the same type of stabilizing coating, it can be noted that there is a tendency for the effective energy $\hbar\omega$ to increase with increasing nanocrystal size. Also, the growth of Fan parameter $A_F$ and the Huang-Rhys factor $S$ testifies an enhance in the exciton-phonon interaction. Similarly, as the QDs swell in size, the temperature coefficients $\beta$ and $\beta_\infty$ go up as well.

*3.3 Inhomogeneous broadening of the exciton absorption band*

The influence of temperature on the OA spectra of condensed matter can also makes itself felt in a change in the half-width of the characteristic bands. Figure 7 outlines the normalized OA spectra of the studied ensembles of nanocrystals at different temperatures. It can be seen that as $T$ drops, the low-energy edge of the exciton band shifts towards higher energies without changing the slope. This behavior indicates a temperature-independent shape of the band.

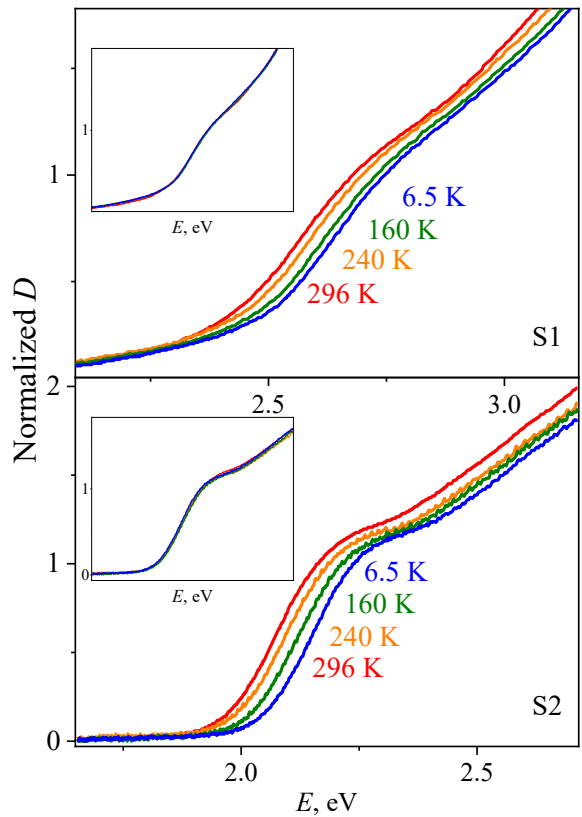
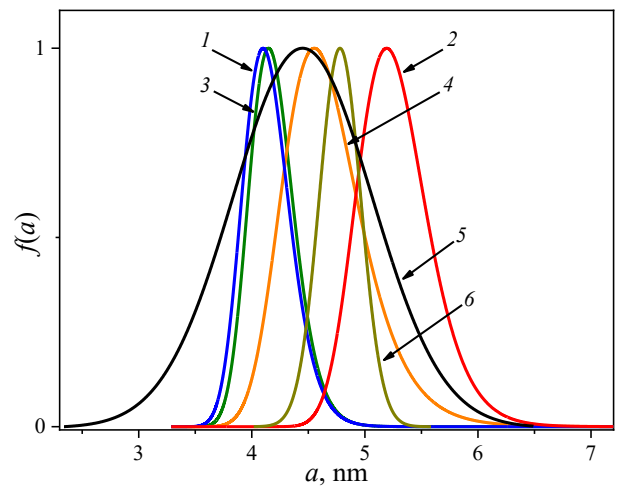

**Figure 7.** The shape of the exciton absorption band of the samples at different $T$.

**Figure 8.** Size distribution for InP/ZnS: 1 – 11.7% (S1), 2 – 13.7% (S2), 3 – 11.1% [43], 4 – 17.4% [43], 5 – 32.6% [24], 6 – 9% [44].

The half-width $H$ of the bands was determined by approximating the low-energy segment of the spectrum by a Gaussian centered at $E_1$ [27]. The values obtained remain unchangeable with varying temperature and are $315 \pm 11$ meV and $227 \pm 5$ meV for the samples S1 and S2, respectively. The observable behavior agrees with the findings for the first exciton absorption band of InP/ZnS, which were analyzed in detail previously [43]. According to the model proposed above, the processes of inhomogeneous broadening caused by a high degree of structural disorder in the ensemble of QDs result in the *H*-temperature independence. This structural disorder is mainly tied up with the size distribution *f(a)* of the nanocrystals. These functions are given in Figure 8. The caption specifies the values of the relative width parameter δ to characterize the quality of the synthesized samples [43].

### 3.4 Temperature dependence of the emission spectra of QDs

The temperature behavior of the luminescence spectra in QDs at hand largely comport with the previously obtained results both for other synthesis methods [42,45–48] and for analogous nanocrystals with a different stabilizer [49–51] (see Table 2). At room temperature, the high-energy $E_x$ component of PL holds for all the cases, whereas, at T < 100 K, the low-energy $E_d$ band makes a significant contribution to the spectrum, see Figure 4. The former is induced by exciton transitions in the core of nanocrystals, the latter arises due to defect states: $Zn_{In}$ substitution atoms in the core, as well as dangling bonds of indium $DB_{In}$ and phosphorus $DB_P$ atoms at the interface [52,53].

**Table 2.** Spectral characteristics of PL in InP/ZnS nanocrystals.

| Size, nm | $E_x$, eV | $H_x$, meV | $E_d$, eV | T range, K | Precursors, injection temperature, coating, manufacturer | Reference |
|---|---|---|---|---|---|---|
| 2.0 (S1) | 2.44– 2.40 | 295 | 2.00 | 6.5–296 | $InX_3$ (X = Cl, Br, I), tris(diethylamino)phosphine, 180 °C, PTVP, NIIPA, Russia | This work |
| 2.6 (S2) | 2.03– 1.97 | 260 | 1.62 | | | |
| 2.1 | 2.36– 2.32 | 231 | 2.04 | 6.5–296 | $InX_3$ (X = Cl, Br, I), tris(diethylamino)phosphine, 180 °C, PAA, NIIPA, Russia | [49] |
| 2.3 | 2.16– 2.13 | 315 | 1.80 | | | |
| 1.8 | 2.45– 2.35 | 165– 236 | – | 2–300 | $In(Ac)_3$, tris(trimethylsilyl)phosphine, 188 °C, myristic acid | [42] |
| 3.0 | 2.14– 2.05 | 134– 217 | – | | | |
| – | 2.05– 2.00 | 192 | 1.80 | 15–300 | $In(Ac)_3$, $PH_3$; 290 °C, myristic acid | [47] |
| 2.9 | 2.02– 1.95 | 166 | 1.66 | 4–290 | $InCl_3$, tris(dimethylamino)phosphine, 160 °C, trioctylphosphine oxide | [48] |
| 2.9 | 2.08–2.06 | 229 | 1.78 | 11–300 | Mesolight Inc., China | [45] |
| 2 | 2.38– 2.35 | 316 | 2.00 | 80–300 | oleic acid, Janus New-Materials Co. Ltd., China | [46] |
| 2.4 | 2.18– 2.15 | 257 | 1.78 | | | |
| 2.9 | 1.84– 1.83 | 310 | 1.66 | | | |
| 4.7 | 1.79– 1.77 | 372 | – | | | |

The half-width $H_x$ of the exciton band is intact over the entire temperature range without exceeding the appropriate values for the half-width of the first exciton band in the OA spectra of the samples. Based on the data presented in the table, it can be inferred that, whatever the synthesis features are, most InP/ZnS quantum dots are characterized by inhomogeneous band broadening processes in the photoluminescence spectra as well.

Figure 9 shows the temperature shift of the exciton emission maximum $E_x$. The solid lines are the approximation of experimental curves by expression (1). The parameter values predicted are listed in Table 3 and generally consistent with the literature data. The effective phonon energy $\hbar\omega$ rests in the range of 9–16 meV and corresponds to the energy of longitudinal acoustic oscillations of 10.2 meV at the L point of the Brillouin zone for bulk InP [54], as well as to the results of our estimation according to the data of [39] (see Table 1). The findings confirm that the shift of the $E_x$ band in InP/ZnS is due to the exciton-phonon interaction with the longitudinal modes of acoustic vibrations.

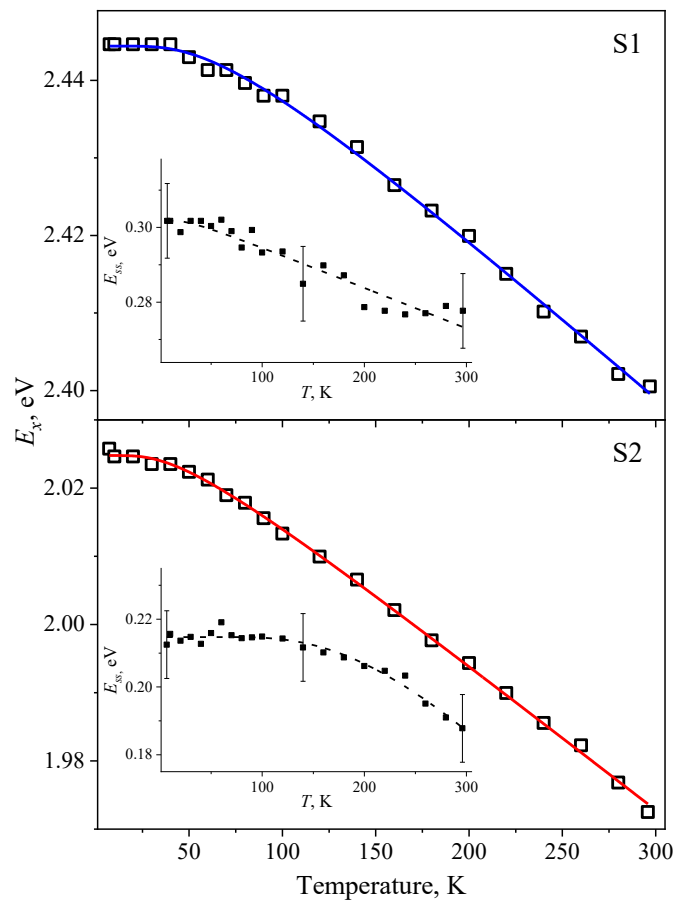

**Figure 9.** Temperature dependence of the energy of the exciton emission maximum of QDs.

Comprehensively analyzing the behavior of the absorption and emission spectra of QDs enables one to trace the temperature dependence of the Stokes shift $E_{ss} = E_1 - E_x$ (see the insets in Figure 9). For the samples studied, as well as for stabilized PAA ones, the value of $E_{ss}$ does not change at low temperatures and declines with further rise of the temperature. The Stokes shift temperature dependence of QDs can be underlain by a fine structure of excited exciton levels [49,55]. The change in $E_{ss}$ within the explored temperature range for the samples S1 and S2 is 30 and 20 meV, respectively, which is 2–3 times less than that for PAA-coated nanocrystals. This fact confirms the influence of the stabilizing shell on the optical properties of InP/ZnS.

**Table 3.** Parameters of the temperature shift of the PL exciton band in QDs.

| Size, nm | $E_x(0)$, ± 0.01 eV | $A_F$, eV | $\hbar\omega$, meV | $S$ | Reference |
|---|---|---|---|---|---|
| 2.0 (S1) | 2.44 | 0.04±0.003 | 16±1 | 1.23 | This work |
| 2.6 (S2) | 2.02 | 0.03±0.002 | 11±1 | 1.25 | |
| 2.1 | 2.35 | 0.04±0.005 | 11±1 | 1.73 | [49] |
| 2.3 | 2.16 | 0.01±0.003 | 9±2 | 0.66 | |
| 2.3 | 2.26 | 0.06* | 13±4 | 2.22±0.07 | |
| 2.8 | 2.21 | 0.06* | 14±5 | 2.15±0.06 | [41] |
| 3.2 | 2.14 | 0.05* | 12±4 | 1.95±0.05 | |
| 1.8 | 2.45 | 0.14* | 23.1±0.1 | 3.02±0.02 | |
| 2.1 | 2.42 | 0.11* | 23.0±0.2 | 2.45±0.01 | |
| 3.0 | 2.14 | 0.14* | 23.3±0.1 | 2.14±0.01 | [42] |
| 3.8 | 1.96 | 0.08* | 21.8±0.5 | 1.96±0.02 | |
| 4.5 | 1.83 | 0.07* | 22.8±0.2 | 1.83±0.02 | |

* our estimate based on the aforecited works.

### 3.5 Quenching of exciton and defect-related luminescence

The efficiency of radiative recombination processes for the investigated QDs noticeably declines with temperature. Figure 10 illustrates the temperature curves for the integral intensities $I_x$ and $I_d$ after normalization to their maximum values. In Figure 3, the analyzed regions of the spectra due to exciton and defect-related luminescence are highlighted by shading. The solid lines stand for the data approximation using the model proposed in [49] and taking into account the distribution of the quenching activation energy $f(E_q)$:

$$I(T) = I_0 \int_0^E \eta(T, E_q) f(E_q) dE_q,$$

$$\eta(T, E_q) = \left[1 + p\exp\left(-\frac{E_q}{kT}\right)\right]^{-1}, \quad (2)$$

$$f(E_q) = (4\ln 2)^{\frac{1}{2}} \pi^{-\frac{1}{2}} H_q^{-1} e^{\left(-4\ln 2(E_q - E_{qm})^2 H_q^{-2}\right)}.$$

Here, the function $\eta(T, E_q)$ governs the efficiency of radiative transitions; $I_0$ is the emission intensity without quenching, a. u.; $p$ is a dimensionless pre-exponential factor that is responsible for the ratio of nonradiative and radiative relaxation rates; $E_q$ is the quenching activation energy, eV. The Gaussian-shaped distribution function has a maximum energy $E_{qm}$ and half-width $H_q$. The calculated values of the model quantities are presented in Table 4. It can be seen that the $f(E_q)$ distribution parameters computed for defect-related emission of the samples are in good agreement with the data for PAA-coated QDs [49] (see Fig. 11). This indicates a general mechanism of quenching of the considered luminescence in the InP/ZnS nanocrystals tested. This mechanism is mainly associated with processes near the core/shell interface and induced by thermally activated transitions of holes from the levels of DB$_P$ dangling phosphorus bonds into the valence band.

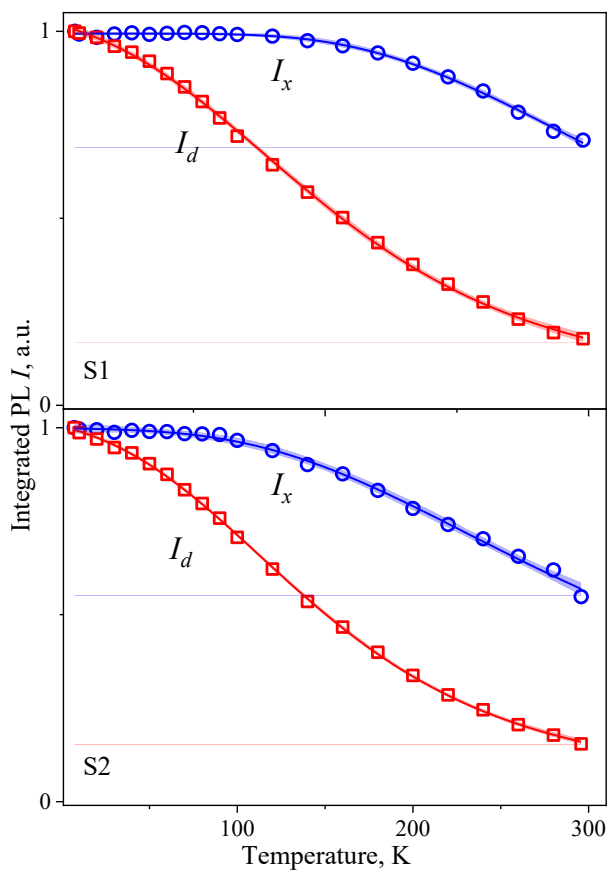

**Figure 10.** Temperature dependence of InP/ZnS PL intensity.

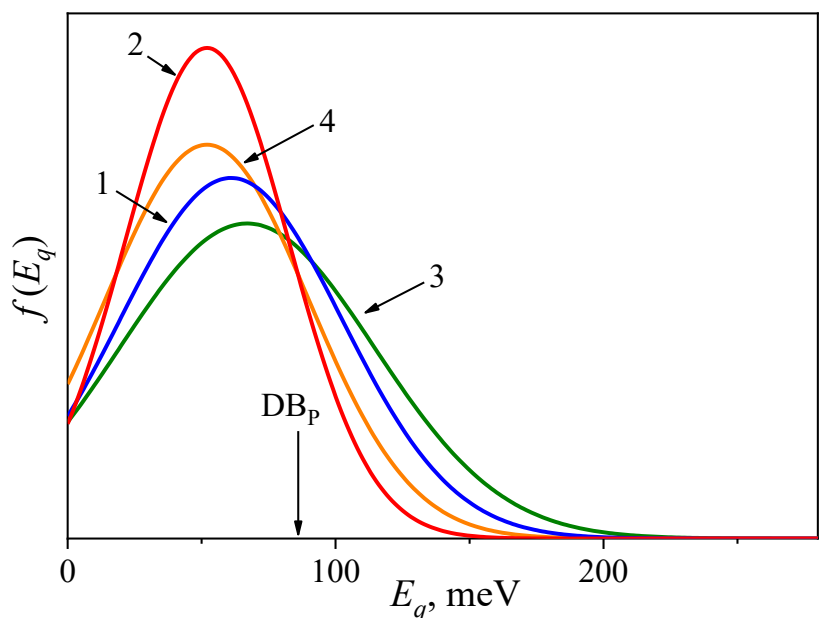

**Figure 11.** Model distribution functions of the quenching activation energy of defect-bound InP/ZnS PL: 1—S1 (2.0 nm), 2—S2 (2.6 nm), 3—InP/ZnS (2.1 nm) [49], 4—InP/ZnS (2.3 nm) [49]. Arrow is the theoretical estimate of the DB$_P$ level depth [53].

**Table 4.** Model parameters of temperature quenching for exciton and defect-related luminescence in InP/ZnS.

| Sample | Emission | $E_{qm}$, ± 10 meV | $H_q$, ±20 meV | $p$ | Quality factor δ, ± 1.0 % |
|---|---|---|---|---|---|
| S1 | exciton | 108 | 54 | 25 | 2.1 quenching<br>12.1 emission<br>11.5 absorption<br>11.7 size |
| | defect-related | 54 | 83 | 69 | – |
| S2 | exciton | 86 | 71 | 21 | 3.3 quenching<br>12.8 emission<br>10.1 absorption<br>13.7 size |
| | defect-related | 52 | 74 | 67 | – |
| InP/ZnS [49] | exciton | 220 | 310 | $2.49 \cdot 10^6$ | 12.1 quenching<br>9.8 emission<br>10.7 absorption<br>11.1 size |
| | defect-related | 69 | 120 | $1.13 \cdot 10^3$ | – |
| InP/ZnS [49] | exciton | 244 | 376 | $6.22 \cdot 10^6$ | 15.6 quenching<br>14.6 emission<br>15.1 absorption<br>17.4 size |
| | defect-related | 52 | 90 | $1.61 \cdot 10^2$ | – |

For quenching of the exciton luminescence of the samples S1 and S2, a lower activation energy is peculiar, as compared with that for InP/ZnS QDs previously studied [49]. This fact can be explained within the proposed mechanism for the escape of an electron from the core into the shell through the thermally activated barrier D$E_{\text{InP/ZnS}}$ [49]. The latter is quantified by the energy difference for the levels corresponding to the bottom of the ZnS and InP conduction bands. An increase in the shell thickness must lead to an increase in the electron affinity energy and, consequently, to a decrease in the activation energy of the quenching barrier. This hypothesis is confirmed both by the results of the studies of exciton PL quenching and by the data obtained from the analysis of the InP/ZnS OA spectra.

## 4. Materials and Methods

Nanocrystals were synthesized at the Research Institute of Applied Acoustics according to a procedure based on the reaction between indium halide and aminophosphine [56]. A heterobifunctional polymer based on polyvinylpyrrolidone-co-maleic anhydride-co-ethylene glycol dimethacrylate with terminal thiol groups (referred to as PTVP) served as a coating. It provides water solubility, colloidal stability, biocompatibility, and simple conjugation of QDs with proteins [57]. The given nanocrystal sizes were calculated on the position of the exciton absorption band of the core using a known relation derived in the effective-mass approximation [4].

Measurements of the optical absorption (OA) spectra of colloidal solutions with different concentrations $c$ of QDs in the range of 0.00005–100 g/L were taken at room temperature inside a quartz cell with an optical path length of 1 cm using a Shimadzu UV-2450 spectrophotometer. For performing optical measurements at below room temperatures, the samples were drop-casted as films onto 1 mm thick quartz substrates. The transmission of the substrates was 94% in the range of 300–900 nm and gradually decreased to 85% at 190 nm. Under the experimental conditions, the substrates exhibited no own luminescent response. The sample temperature was controlled using a CCS-100/204N closed-loop helium cryostat (Janis Research Company, LLC). The measurements were carried out at the following temperatures: $T = 6.5$ K, in the range of 10–100 K with a step of 10 K, from 100 to 296 K every 20 K.

To explore the temperature-dependence of the OA spectra, we assembled a measuring setup equipped with a spectrophotometer and a cryogenic system [27,58]. The measurements were carried out in the wavelength range from 190 to 900 nm at a scanning rate of 225 nm/min. The scanning step and spectral width of the slit were 0.5 and 2 nm, respectively. Photoluminescence (PL) spectra were analyzed using a Shamrock SR-303i-B spectrograph and a Newton[EM] DU-970P-BV-602 CCD (Andor Technology) with cooling up to - 80°C. The samples were excited by a LED with a maximum at 372 nm and an average power density of 5 mW/cm$^2$. The luminescence quantum yield of nanocrystals at room temperature was estimated using a Rhodamine 6G [59] and was 6% and 17% for S1 and S2, respectively.

## 5. Conclusions

A comparative analysis of the spectral features of optical absorption and photoluminescence in the temperature range of 6.5–296 K was performed for QD samples stabilized with a heterobifunctional polyvinylpyrrolidone-based polymer (PTVP). It is shown that the type of coating affects the interaction between individual nanocrystals when depositng from a solution. It is established that the temperature shifts of the positions of the first exciton absorption band $E_1$ and exciton emission $E_x$ are due chiefly to the interaction with the modes of acoustic vibrations having an effective energy of 9–23 meV. The half-width $H$ of the first exciton absorption band is found to be unchanged in the temperature range under study. This fact agrees with the data on PAA-stabilized nanocrystals and is controlled by the wide size distribution of QDs in the ensemble. Whatever the synthesis technique and coating are, the temperature behavior of the exciton emission of the InP/ZnS ensembles evidences an inhomogeneous broadening of the spectra. The observable temperature dependence of the Stokes shift can be caused by a fine structure of the excited exciton levels. It is proved that the quenching of defect-related luminescence occurs involving mainly the energy levels of DB$_P$ dangling phosphorus bonds at the core-shell interface. In turn, the activation energy of exciton emission quenching depends on the thickness of the ZnS shell. This statement gives a right to claim that the proposed mechanism provides non-radiative relaxation due to the thermal escape of an excited electron beyond the InP core.

**Funding:** The research is supported by Minjbrnauki of the Russian Federation, project FEUZ-2023-0014.

**References**

1. Efros, Al.L.; Efros, A.L. Interband Absorption of Light in a Semiconductor Sphere. *Sov. Phys. Semicond.*, 1982, **16**, 772–775.

2. Brus, L.E. Electron-Electron and Electron-Hole Interactions in Small Semiconductor Crystallites: The Size Dependence of the Lowest Excited Electronic State. *J. Chem. Phys.* **1984**, *80*, 4403–4409, doi:10.1063/1.447218.

3. Gaponenko, S.V. Optical processes in semiconductor nanocrystallites (quantum dots). *Semicond.,* 1996, **30**, 315–336.

4. Gaponenko, S. V. *Optical Properties of Semiconductor Nanocrystals*; Cambridge University Press, 1998; ISBN 9780521582414.

5. Alivisatos, A.P. Semiconductor Clusters, Nanocrystals, and Quantum Dots. *Science (80-. ).* **1996**, *271*, 933–937, doi:10.1126/science.271.5251.933.

6. Efros, A.L.; Brus, L.E. Nanocrystal Quantum Dots: From Discovery to Modern Development. *ACS Nano* **2021**, *15*, 6192–6210, doi:10.1021/acsnano.1c01399.

7. Podshivaylov, E.A.; Kniazeva, M.A.; Gorshelev, A.A.; Eremchev, I.Y.; Naumov, A. V.; Frantsuzov, P.A. Contribution of Electron-Phonon Coupling to the Luminescence Spectra of Single Colloidal Quantum Dots. *J. Chem. Phys.* **2019**, *151*, 174710, doi:10.1063/1.5124913.

8. Savchenko, S.S.; Vokhmintsev, A.S.; Weinstein, I.A. Luminescence Parameters of InP/ZnS@AAO Nanostructures. *AIP Conf. Proc.* **2016**, *1717*, 040028, doi:10.1063/1.4943471.

9. Savchenko, S.S.; Vokhmintsev, A.S.; Weinstein, I.A. Photoluminescence Thermal Quenching of Yellow-Emitting InP/ZnS Quantum Dots. In Proceedings of the AIP Conference Proceedings; 2018; Vol. 2015, p. 020085.

10. Anc, M.J.; Pickett, N.L.; Gresty, N.C.; Harris, J.A.; Mishra, K.C. Progress in Non-Cd Quantum Dot Development for Lighting Applications. *ECS J. Solid State Sci. Technol.* **2013**, *2*, R3071–R3082, doi:10.1149/2.016302jss.

11. Heath, J.R.; Shiang, J.J. Covalency in Semiconductor Quantum Dots. *Chem. Soc. Rev.* **1998**, *27*, 65–71, doi:10.1039/a827065z.

12. Brichkin, S.B. Synthesis and properties of colloidal indium phosphide quantum dots. *Colloid J.*, 2015, **77**, 393–403; DOI: 10.1134/S1061933X15040043.

13. Mushonga, P.; Onani, M.O.; Madiehe, A.M.; Meyer, M. Indium Phosphide-Based Semiconductor Nanocrystals and Their Applications. *J. Nanomater.* **2012**, *2012*, 869284, doi:10.1155/2012/869284.

14. Brunetti, V.; Chibli, H.; Fiammengo, R.; Galeone, A.; Malvindi, M.A.; Vecchio, G.; Cingolani, R.; Nadeau, J.L.; Pompa, P.P. InP/ZnS as a Safer Alternative to CdSe/ZnS Core/Shell Quantum Dots: In Vitro and in Vivo Toxicity Assessment. *Nanoscale* **2013**, *5*, 307–317, doi:10.1039/C2NR33024E.

15. Chibli, H.; Carlini, L.; Park, S.; Dimitrijevic, N.M.; Nadeau, J.L. Cytotoxicity of InP/ZnS Quantum Dots Related to Reactive Oxygen Species Generation. *Nanoscale* **2011**, *3*, 2552, doi:10.1039/c1nr10131e.

16. Ayupova, D.; Dobhal, G.; Laufersky, G.; Nann, T.; Goreham, R. An In Vitro Investigation of Cytotoxic Effects of InP/Zns Quantum Dots with Different Surface Chemistries. *Nanomaterials* **2019**, *9*, 135, doi:10.3390/nano9020135.

17. Ayed, Z.; Malhotra, S.; Dobhal, G.; Goreham, R. V. Aptamer Conjugated Indium Phosphide Quantum Dots with a Zinc Sulphide Shell as Photoluminescent Labels for Acinetobacter Baumannii. *Nanomaterials* **2021**, *11*, 3317, doi:10.3390/nano11123317.

18. Yong, K.T.; Ding, H.; Roy, I.; Law, W.C.; Bergey, E.J.; Maitra, A.; Prasad, P.N. Imaging Pancreatic Cancer Using Bioconjugated Inp Quantum Dots. *ACS Nano* **2009**, *3*, 502–510, doi:10.1021/nn8008933.

19. Massadeh, S.; Nann, T. InP/ZnS Nanocrystals as Fluorescent Probes for the Detection of ATP. *Nanomater. Nanotechnol.* **2014**, *4*, 1–8, doi:10.5772/58523.

20. Zhang, J.; Wang, Y.; Xu, Z.; Shi, C.; Yang, X. A Sensitive Fluorescence-Visualized Sensor Based on an InP/ZnS Quantum Dots-Sodium Rhodizonate System for Monitoring Fish Freshness. *Food Chem.* **2022**, *384*, 132521, doi:10.1016/j.foodchem.2022.132521.

21. Li, C.; Ando, M.; Enomoto, H.; Murase, N. Highly Luminescent Water-Soluble InP/ZnS Nanocrystals Prepared via Reactive Phase Transfer and Photochemical Processing. *J. Phys. Chem. C* **2008**, *112*, 20190–20199, doi:10.1021/jp805491b.

22. Bang, E.; Choi, Y.; Cho, J.; Suh, Y.-H.; Ban, H.W.; Son, J.S.; Park, J. Large-Scale Synthesis of Highly Luminescent InP@ZnS Quantum Dots Using Elemental Phosphorus Precursor. *Chem. Mater.* **2017**, *29*, 4236–4243, doi:10.1021/acs.chemmater.7b00254.

23. Huang, F.; Bi, C.; Guo, R.; Zheng, C.; Ning, J.; Tian, J. Synthesis of Colloidal Blue-Emitting InP/ZnS Core/Shell Quantum Dots with the Assistance of Copper Cations. *J. Phys. Chem. Lett.* **2019**, *10*, 6720–6726, doi:10.1021/acs.jpclett.9b02386.

24. Arias-Ceron, J.S.; Gonzalez-Araoz, M.P.; Bautista-Hernandez, A.; Sanchez Ramirez, J.F.; Herrera-Perez, J.L.; J.G., M.-A. Semiconductor Nanocrystals of InP @ ZnS : Synthesis and Characterization. *Superf. y Vacío* **2012**, *25*, 134–138.

25. Talsky, G. *Derivative Spectrophotometry: Low and Higher Order*; VCH: Weinheim, 1994; ISBN 3527282947.

26. Savchenko, S.; Vokhmintsev, A.; Weinstein, I. Exciton–Phonon Interactions and Temperature Behavior of Optical Spectra in Core/Shell InP/ZnS Quantum Dots. In *Core/Shell Quantum Dots: Synthesis, Properties and Devices*; Tong, X., Wang, Z.M., Eds.; Springer: Cham, 2020; pp. 165–196 ISBN 9783030465957.

27. Savchenko, S.S.; Vokhmintsev, A.S.; Weinstein, I.A. Temperature-Induced Shift of the Exciton Absorption Band in InP/ZnS Quantum Dots. *Opt. Mater. Express* **2017**, *7*, 354–359, doi:10.1364/OME.7.000354.

28. Savchenko, S.S.; Vokhmintsev, A.S.; Weinstein, I.A. Effect of Temperature on the Spectral Properties of InP/ZnS Nanocrystals. *J. Phys. Conf. Ser.* **2018**, *961*, 012003, doi:10.1088/1742-6596/961/1/012003.

29. Mićić, O.I.; Jones, K.M.; Cahill, A.; Nozik, A.J. Optical, Electronic, and Structural Properties of Uncoupled and Close-Packed Arrays of InP Quantum Dots. *J. Phys. Chem. B* **1998**, *102*, 9791–9796, doi:10.1021/jp981703u.

30. Fu, H.; Zunger, A. Excitons in InP Quantum Dots. *Phys. Rev. B* **1998**, *57*, R15064–R15067, doi:10.1103/PhysRevB.57.R15064.

31. Savchenko, S.S.; Vokhmintsev, A.S.; Weinstein, I.A. Temperature dependence of the optical absorption spectra of InP/ZnS quantum dots. *Tech. Phys. Lett.*, 2017, **43**, 297–300; DOI: 10.1134/S1063785017030221.

32. A.P. Babichev, *Handbook of Physical Quantities*, Ed. I.S. Grigor'ev, E.Z. Meilikhov, Energoatomizdat, Moscow, 1991, 1232 p. (in Russian).

33. Weller, H.; Koch, U.; Gutiérrez, M.; Henglein, A. Photochemistry of Colloidal Metal Sulfides. 7. Absorption and Fluorescence of Extremely Small ZnS Particles (The World of the Neglected Dimensions). *Berichte der Bunsengesellschaft für Phys. Chemie* **1984**, *88*, 649–656, doi:10.1002/bbpc.19840880715.

34. Zhang, Y.; Ma, M.; Wang, X.; Fu, D.; Gu, N.; Liu, J.; Lu, Z.; Ma, Y.; Xu, L.; Chen, K. First-Order Hyperpolarizability of ZnS Nanocrystal Quantum Dots Studied by Hyper-Rayleigh Scattering. *J. Phys. Chem. Solids* **2002**, *63*, 2115–2118, doi:10.1016/S0022-3697(02)00259-7.

35. Kho, R.; Torres-Martínez, C.L.; Mehra, R.K. A Simple Colloidal Synthesis for Gram-Quantity Production of Water-Soluble ZnS Nanocrystal Powders. *J. Colloid Interface Sci.* **2000**, *227*, 561–566, doi:10.1006/jcis.2000.6894.

36. Savchenko, S.S.; Vokhmintsev, A.S.; Weinstein, I.A. Optical Properties of InP/ZnS Quantum Dots Deposited into Nanoporous Anodic Alumina. *J. Phys. Conf. Ser.* **2016**, *741*, 012151, doi:10.1088/1742-6596/741/1/012151.

37. Vainshtein, I.A.; Zatsepin, A.F.; Kortov, V.S. Applicability of the empirical Varshni relation for the temperature dependence of the width of the band gap. *Phys. Solid State.*, 1999, **41**, 905–908; DOI: 10.1134/1.1130901.

38. Karimullin, K.R.; Arzhanov, A.I.; Eremchev, I.Y.; Kulnitskiy, B.A.; Surovtsev, N. V.; Naumov, A. V. Combined Photon-Echo, Luminescence and Raman Spectroscopies of Layered Ensembles of Colloidal Quantum Dots. *Laser Phys.* **2019**, *29*, 124009, doi:10.1088/1555-6611/ab4bdb.

39. Turner, W.J.; Reese, W.E.; Pettit, G.D. Exciton Absorption and Emission in InP. *Phys. Rev.* **1964**, *136*, 1955–1958, doi:10.1103/PhysRev.136.A1467.

40. Zilli, A.; De Luca, M.; Tedeschi, D.; Fonseka, H.A.; Miriametro, A.; Tan, H.H.; Jagadish, C.; Capizzi, M.; Polimeni, A. Temperature Dependence of Interband Transitions in Wurtzite InP Nanowires. *ACS Nano* **2015**, *9*, 4277–4287, doi:10.1021/acsnano.5b00699.

41. Narayanaswamy, A.; Feiner, L.F.; van der Zaag, P.J. Temperature Dependence of the Photoluminescence of InP/ZnS Quantum Dots. *J. Phys. Chem. C* **2008**, *112*, 6775–6780, doi:10.1021/jp800339m.

42. Narayanaswamy, A.; Feiner, L.F.; Meijerink, A.; van der Zaag, P.J. The Effect of Temperature and Dot Size on the Spectral Properties of Colloidal InP/ZnS Core−Shell Quantum Dots. *ACS Nano* **2009**, *3*, 2539–2546, doi:10.1021/nn9004507.

43. Savchenko, S.S.; Weinstein, I.A. Inhomogeneous Broadening of the Exciton Band in Optical Absorption Spectra of InP/ZnS Nanocrystals. *Nanomaterials* **2019**, *9*, 716, doi:10.3390/nano9050716.

44. Shen, W.; Tang, H.; Yang, X.; Cao, Z.; Cheng, T.; Wang, X.Y.; Tan, Z.; Deng, Z.; You, J. Synthesis of Highly Fluorescent InP/ZnS Small-Core/Thick-Shell Tetrahedral-Shaped Quantum Dots for Blue Light-Emitting Diodes. *J. Mater. Chem. C* **2017**, *5*, 8243–8249, doi:10.1039/C7TC02927F.

45. Chen, T.; Li, K.; Mao, H.; Chen, Y.; Wang, J.; Weng, G. Photoluminescence Investigation of the InP/ZnS Quantum Dots and Their Coupling with the Au Nanorods. *J. Electron. Mater.* **2019**, *48*, 3497–3503, doi:10.1007/s11664-019-07106-9.

46. Wang, C.; Wang, Q.; Zhou, Z.; Wu, W.; Chai, Z.; Gao, Y.; Kong, D. Temperature Dependence of Photoluminescence Properties in InP/ZnS Core-Shell Quantum Dots. *J. Lumin.* **2020**, *225*, 117354, doi:10.1016/j.jlumin.2020.117354.

47. Pham, T.T.; Chi Tran, T.K.; Nguyen, Q.L. Temperature-Dependent Photoluminescence Study of InP/ZnS Quantum Dots. *Adv. Nat. Sci. Nanosci. Nanotechnol.* **2011**, *2*, 025001, doi:10.1088/2043-6262/2/2/025001.

48. Biadala, L.; Siebers, B.; Beyazit, Y.; Tessier, M.D.; Dupont, D.; Hens, Z.; Yakovlev, D.R.; Bayer, M. Band-Edge Exciton Fine Structure and Recombination Dynamics in InP/ZnS Colloidal Nanocrystals. *ACS Nano* **2016**, *10*, 3356–3364, doi:10.1021/acsnano.5b07065.

49. Savchenko, S.S.; Vokhmintsev, A.S.; Weinstein, I.A. Activation Energy Distribution in Thermal Quenching of Exciton and Defect-Related Photoluminescence of InP/ZnS Quantum Dots. *J. Lumin.* **2022**, *242*, 118550, doi:10.1016/j.jlumin.2021.118550.

50. Savchenko, S.S.; Vokhmintsev, A.S.; Weinstein, I.A. Non-Radiative Relaxation Processes in Luminescence of InP/ZnS Quantum Dots. *J. Phys. Conf. Ser.* **2020**, *1537*, 012015, doi:10.1088/1742-6596/1537/1/012015.

51. Savchenko, S.S.; Vokhmintsev, A.S.; Weinstein, I.A. Spectral Features and Luminescence Thermal Quenching of InP/ZnS Quantum Dots within 7.5 – 295 K Range. In Proceedings of the Advanced Photonics 2018 (BGPP, IPR, NP, NOMA, Sensors, Networks, SPPCom, SOF); OSA: Washington, D.C., 2018; Vol. Part F107-, p. NoW1J.4.

52. Janke, E.M.; Williams, N.E.; She, C.; Zherebetskyy, D.; Hudson, M.H.; Wang, L.; Gosztola, D.J.; Schaller, R.D.; Lee, B.; Sun, C.; et al. Origin of Broad Emission Spectra in InP Quantum Dots: Contributions from Structural and Electronic Disorder. *J. Am. Chem. Soc.* **2018**, *140*, 15791–15803, doi:10.1021/jacs.8b08753.

53. Cho, E.; Kim, T.; Choi, S.; Jang, H.; Min, K.; Jang, E. Optical Characteristics of the Surface Defects in InP Colloidal Quantum Dots for Highly Efficient Light-Emitting Applications. *ACS Appl. Nano Mater.* **2018**, *1*, 7106–7114, doi:10.1021/acsanm.8b01947.

54. Alfrey, G.F.; Borcherds, P.H. Phonon Frequencies from the Raman Spectrum of Indium Phosphide. *J. Phys. C Solid State Phys.* **1972**, *5*, L275–L278, doi:10.1088/0022-3719/5/20/002.

55. Watanabe, T.; Takahashi, K.; Shimura, K.; Kim, D. Influence of Carrier Localization at the Core/Shell Interface on the Temperature Dependence of the Stokes Shift and the Photoluminescence Decay Time in CdTe/CdS Type-II Quantum Dots. *Phys. Rev. B* **2017**, *96*, 035305, doi:10.1103/PhysRevB.96.035305.

56. Tessier, M.D.; Dupont, D.; De Nolf, K.; De Roo, J.; Hens, Z. Economic and Size-Tunable Synthesis of InP/ZnE (E = S, Se) Colloidal Quantum Dots. *Chem. Mater.* **2015**, *27*, 4893–4898, doi:10.1021/acs.chemmater.5b02138.

57. Dezhurov, S. V.; Krylsky, D. V.; Rybakova, A. V.; Ibragimova, S.A.; Gladyshev, P.P.; Vasiliev, A.A.; Morenkov, O.S. One-Pot Synthesis of Polythiol Ligand for Highly Bright and Stable

Hydrophilic Quantum Dots toward Bioconjugate Formation. *Adv. Nat. Sci. Nanosci. Nanotechnol.* **2018**, *9*, 015002, doi:10.1088/2043-6254/aa9de8.

58. Вохминцев, А.С.; Минин, М.Г.; Чайкин, Д.В.; Вайнштейн, И.А. Высокотемпературная Приставка Для Измерения Спектральных Характеристик Термолюминесценции. *Приборы и техника эксперимента* **2014**, *2014*, 139–143, doi:10.7868/S0032816214020323.

59. Grabolle, M.; Spieles, M.; Lesnyak, V.; Gaponik, N.; Eychmüller, A.; Resch-Genger, U. Determination of the Fluorescence Quantum Yield of Quantum Dots: Suitable Procedures and Achievable Uncertainties. *Anal. Chem.* **2009**, *81*, 6285–6294, doi:10.1021/ac900308v.